\documentclass[preprint]{aastex}
\usepackage{graphicx,xfrac,amsmath,longtable,color}

\def\la{\mathrel{\hbox{\rlap{\hbox{\lower4pt\hbox{$\sim$}}}\hbox{$<$}}}}
\def\ga{\mathrel{\hbox{\rlap{\hbox{\lower4pt\hbox{$\sim$}}}\hbox{$>$}}}}

\def\dm15{{$\Delta$}$m_{15}$}

\def\v10{$V_{10}$(Si~II)}

\def\W575{$W(5750)$}
\def\W610{$W(6100)$}
\def\6100{the 6100~\AA\ absorption}

\def\msun{~M$_\odot$}

\def\CaII7291{Ca {\sc II}] $\lambda\lambda$ 7291,7323\ }

\def\OI6300{[O {\sc I}] $\lambda\lambda$ 6300,6364\ }

\def\apj{Ap. J.}
\def\apjl{Ap. J. Lett.}
\def\apjs{Ap. J. Supp.}

\def\nat{Nature}

\def\aap{Astron. \& Astroph.}

\def\apss{Ap. and Space. Sci.}

\newcommand{\ergsec}{\ensuremath{\mathrm{erg~s}^{-1}}}
\newcommand{\galex}{\textit{GALEX}}

\begin{document}

\title{Self-shielding of Soft X-rays in SN~Ia Progenitors}

\author{J. Craig Wheeler\altaffilmark{1}, D. Pooley\altaffilmark{2}}
\authoremail{wheel@astro.as.utexas.edu}
\altaffiltext{1}{Department of Astronomy, University of Texas at Austin,
Austin, TX, USA.}
\altaffiltext{2}{Department of Physics, Sam Houston State University, Huntsville, TX, USA}

\begin{abstract}

There are insufficient super soft ($\sim 0.1$ keV) X-ray sources in 
either spiral or elliptical galaxies to account for the rate of 
explosion of Type Ia supernovae in either the single degenerate or 
the double degenerate scenarios. We quantify the amount of circumstellar 
matter that would be required to suppress the soft X-ray flux by 
yielding a column density in excess of $10^{23}$ cm$^{-2}$. We summarize
evidence that appropriate quantities of matter are extant in SN~Ia
and in recurrent novae that may be supernova precursors. The obscuring 
matter is likely to have a large, but not complete, covering factor and
to be substantially non-spherically symmetric. Assuming that much of the 
absorbed X-ray flux is re-radiated as black-body radiation in the UV, we
estimate that $\lesssim$ 100 sources might be detectable in the \galex\
all sky survey.

\end{abstract}



\section{Introduction}

The first suggestion that Type Ia (SN~Ia) supernovae may arise 
through mass transfer from a non-degenerate star onto a white
dwarf may have been Wheeler \& Hansen (1971). This suggestion
of a single degenerate (SD) model was quantified by Whelan \& 
Iben (1975) and pursued by many since. An alternative model is 
the merger of two degenerate stars, the double-degenerate (DD)
model (Iben \& Tutukov 1984; Webbink 1984). Both of these
models are constrained by the paucity of bright, soft, X-ray
sources (Di Stefano 2010a,b). In the SD model, an associated
constraint is the strong expectation that the mass transfer 
rate must be sufficiently large that accretion leads to 
non-degenerate shell burning on the surface of the white dwarf 
in order to avoid classical nova explosions that eject the 
accreted matter and, probably, some of the white dwarf material 
as well (see, e.g., Nomoto 1982; Iben 1982; Fujimoto 1982; 
Shen \& Bildsten 2008; and references therein). This constraint 
requires the progenitor to be bright and hot, qualities exhibited 
by the super-soft X-ray sources (SSS; van den Heuvel et al. 1992; 
Kahabka \& van den Heuvel 1997). The problem is that there are 
not enough SSS seen in either spiral or ellliptical galaxies to 
account for the rate of production of SN~Ia by about a factor of 
order 100 (Di Stefano 2010a). Similar constraints arise for the DD
model. Binary synthesis models require that the progenitor 
systems go through a phase of rapid accretion onto the primary 
white dwarf prior to the common envelope phase that reveals the 
second white dwarf. One way to avoid these constraints is to 
shroud the progenitor systems in sufficient material that soft 
X-rays may be produced, but absorbed and transmuted into other 
wavelengths rather than radiated directly. One possibility is 
the production of winds from the surface of the accreting white 
dwarf (Hachisu, Kato \& Nomoto 1992; Kato \& Hachisu 1999;
Hachisu, Kato \& Nomoto 2010). A constraint on this particular 
suggestion is the lack of evidence for such a wind in the remnant 
of SN Ia 1572 (Badenes et al. 2007).  

Here we explore the general constraints on circumstellar 
matter that might produce sufficient absorption to suppress
super-soft ($\sim$ 0.1 keV) X-rays, describe two lines of 
evidence that such circumstellar absorption exists,
discusss the bands in which such absorbed soft flux might
be re-emitted and the liklihood that such systems could 
be observed.  

\section{Soft X-ray Column Densities}

\begin{figure}[htp]
\centering
\includegraphics[width=\textwidth]{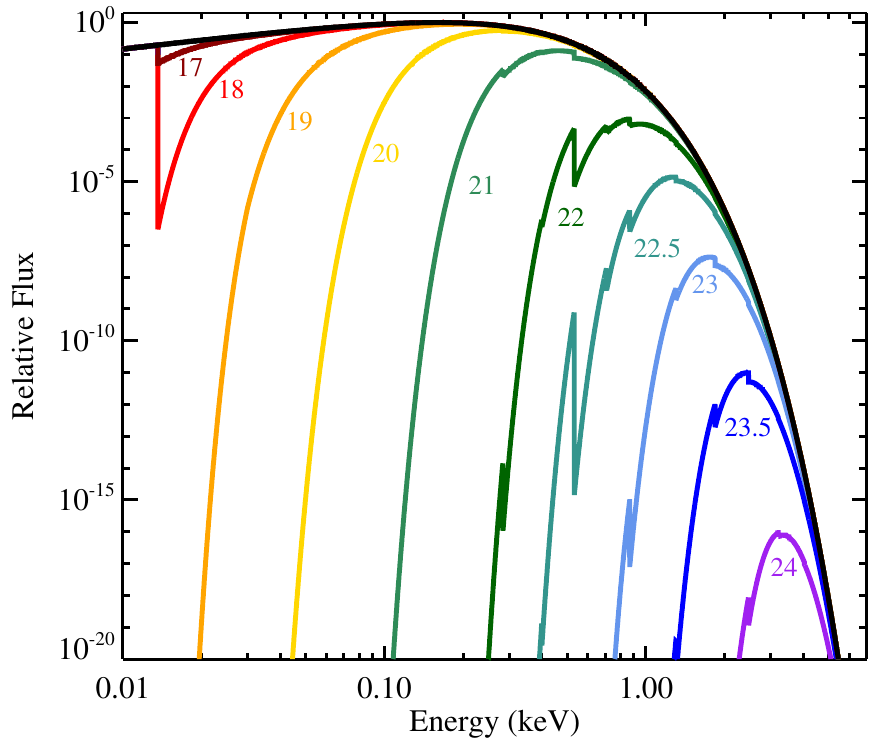}
\figcaption[absorbed_bb.pdf]
{Spectra of a thermal black body of temperature 0.1 keV corresponding
to a super-soft X-ray source subject to absorption by varying
column depths of solar abundance matter.
\label{absorb}}
\end{figure}

\begin{figure}[htp]
\centering
\includegraphics[width=\textwidth]{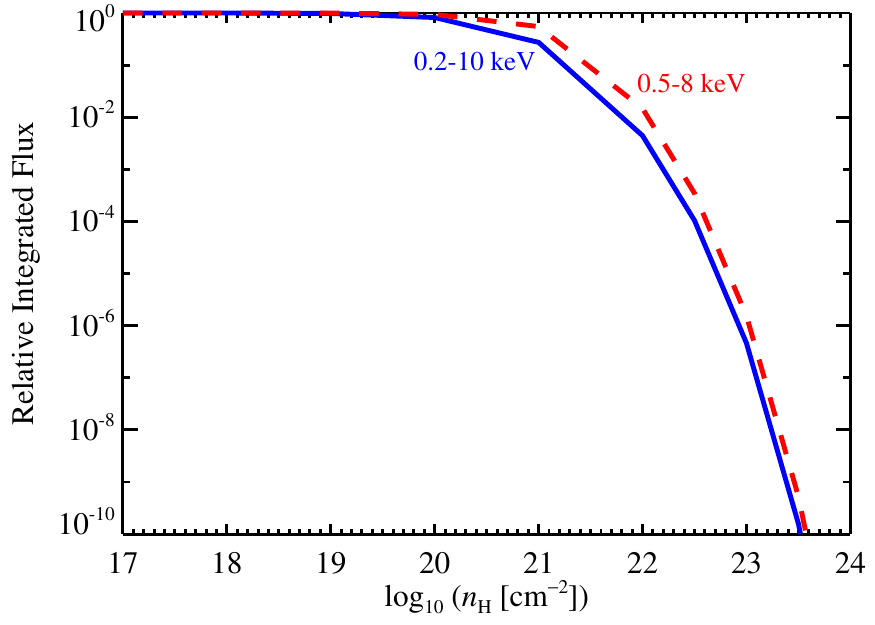}
\figcaption[bb_relflux.pdf]
{The absorbed integrated flux of $kT = 0.1$~keV blackbody relative to the unabsorbed flux as function of column density.  The blue curve shows the 0.2--10 keV band, commonly used for {\it XMM} observations, and the red curve shows the 0.5--8 keV band, commonly used for {\it Chandra} observations.
\label{bbrel}}
\end{figure}

Observations of SSS and models of rapidly accreting, shell-burning 
white dwarfs suggest that the flux emerges from the surface of the 
accreting white dwarf with thermal spectra at energies of about 
0.1 keV (Kahabka \& van den Heuvel 1997). Very little matter is 
required to absorb this flux. Figure \ref{absorb} shows the emergent 
spectrum of a thermal black body with an effective temperature of 
0.1 keV subject to absorption by a range of column depths. 
This figure shows that for such a characteristic thermal emission, 
a column depth of solar abundance matter of $10^{23}$ cm$^{-2}$ 
would reduce the flux density at about 1 keV by a factor of about 100.  
Detection in X-rays is based on the integrated flux over a standard bandpass.  
Figure~\ref{bbrel} shows the effects of absorption on this integrated flux.  
The integrated flux is down by a factor of 100 around a column of $10^{22}$ 
cm$^{-2}$.  For a column density of $10^{23}$ cm$^{-2}$, the integrated flux 
is reduced by a factor of roughly $10^6$. Such column depths would, in principle, 
solve the problem of the paucity of SSS. Any higher column depth would 
effectively totally block such soft flux.

Any circumstellar matter must be rather sparse and dilute in order 
not to perturb the early light curve of SN Ia on the rise to 
maximum (Kasen 2010; Hayden et al. 2010). If there is circumstellar 
matter in a SD configuration, then there is some {\it a priori} 
expectation that it resides at a distance representative of the 
size of the orbit. In a typical mass-transferring situation, that 
would be an orbital period of hours to days or a radius of order
$10^{11}$ to $10^{12}$ cm. To have column depth of 
$\sigma = 10^{23}$ cm$^{-2}$ at this radius requires a particle 
density of $n \sim 10^{11} \sigma_{23} R_{12}^{-1}$ cm$^{-3}$ or 
a mass density of $\rho \sim 1.6\times10^{-13} \sigma_{23} 
R_{12}^{-1}$ g cm$^{-3}$. A density of this order in a volume of 
this radius implies a mass of 
\begin{equation}
\label{mass}
m_{\rm csm} \sim 3\times10^{-10} \sigma_{23} R_{12}^2~M_{\odot}. 
\end{equation}
Minor amounts of matter could easily obscure a SSS. 
Both mass transfer and possible shell burning on the surfaces
of white dwarfs are likely to be messy events. We next explore
the degree to which this ``mess" is likely to produce a CSM
that obscures any soft X-rays.

\section{Absorption in SN~Ia}

While there are indications of circumstellar absorption in some normal 
SN~Ia (Patat et al. 2007; Simon et al. 2009; Blondin et al. 2009)
this behavior is rare. Some SN~Ia-like events show obvious evidence
for substantial circumstellar hydrogen (SN~2001ic; Hamuy et al. 2003), 
but again
these are rare, peculiar events. A more common hint of circumstellar
matter may be revealed in the high-velocity Ca features routinely
seen in early spectra of SN~Ia (Wang et al. 2003; Mazzali et al. 2005).
These features are not seen in every SN~Ia, but may appear in of
order 80\% (Marion, private communication). There is no direct evidence
that these high-velocity features are associated with circumstellar
matter, but a plausible model has been presented by Gerardy et al. (2004)
in which circumstellar matter is impacted by the ejecta of a 
delayed-detonation model of a SN~Ia. The collision leads to the formation 
of a shell at the contact discontinuity. The resulting shell is dense 
enough to reveal strong absorption in the lines of a solar abundance 
of calcium in the form of the Ca II IR triplet where the high-velocity 
feature is most often observed. The models suggests that it would still 
be very difficult to directly detect the hydrogen (or helium) substrate. 
To leave the dense shell at the observed velocity, it must have a mass of     
about 0.02\msun. Gerardy et al. point out that this circumstellar 
matter must be at a radius $<< 10^{15}$ cm, the typical radius for
the photosphere of a supernova at maximum light, so that the energy
of the collision is radiated quickly and does not adversely affect the
early rise of the light curve. 

If we take this model at face value, we can estimate the implied column 
depth to soft X-rays. Taking the mass to be $10^{31} M_{31}$ g at a radius 
of less than $10^{14} R_{14}$ cm will yield a mass density of about $2.4\times
10^{-12} M_{31} R_{14}^{-3}$ g cm$^{-3}$ or a number density of about
$1.4\times 10^{12} M_{31} R_{14}^{-3}$ cm$^{-3}$. At this radius, the
column density would be 
\begin{equation}
\label{HV}
\sigma \sim 1.4\times 10^{26} M_{31} R_{14}^{-2} \rm{cm^{-2}}, 
\end{equation}
a huge column depth. If the putative dense shell were at smaller radius, the column density
would be even larger. Such a shell
of absorbing material would be ample to absorb the soft X-rays and
render an underlying source unobservable to either {\it Chandra} or {\it XMM}.

The high-velocity feature is strongly polarized (Wang et al. 2003;
Wang \& Wheeler 2008; Patat et al. 2009) and hence asymmetric. 
Whatever this feature is, it shows a different perspective from
different aspect angles. The fact that a stong majority of SN~Ia
reveal this feature means that its covering factor is large. Since
some SN~Ia do not show this feature, the covering factor is likely
not to be 100\%. Perhaps 10 - 20\% of the sky is free of sufficient
material to yield the appropriate absorption. This large a percentage
would be too large for the observed suppression of SSS sources, but
as we have illustrated, very little mass is required for the latter. 
X-ray absorbing mass may ``fill the hole" that is implied by the lack 
of complete coverage of the polarized high-velocity feature without
yielding substantial Ca IR triplet absorption on that particular
line of sight.

\section{Recurrent Novae}

Recurrent novae have long been discussed as possible precursor systems
for SN~Ia (Schaefer 2010, and references therein). Their shell burning
is sporadic and the masses of the white dwarf are generally thought to
be large and growing. The recurrent nova eruption is sure to eject
some matter into the circumstellar environment. Patat et al. (2011;
see also Patat 2011) have obtained high-resolution spectroscopy of
the recurrent nova RS Oph before, during, and after its 2006 outburst.
They deduce that the eruptions create complex structures within the
material lost by the donor star. There are signs of the interaction
of matter ejected in the most recent outburst with matter ejected in
the previous outburst in 1986. Kinematics put that interaction at a 
radius less than $4\times10^{14}$ cm, within the volume of the photosphere
of a supernova at maximum light. The evidence suggests that recurrent
novae outbursts do not destroy the slow-moving shells produced in
previous outbursts and that recurrent novae are able to produce
long-lasting structures in their circumstellar environments. Patat
et al. emphasize the similarity of the Na D absorption structure
in RS Oph compared to a sample of SN~Ia. The mass involved in this
circumstellar matter is not well constrained, but Patat et al. 
estimate masses of order $10^{-5}$\msun, presumed to arise in 
a wind from a red giant companion. With this mass, we can estimate
a column density of 
\begin{equation}
\label{RN}
\sigma \sim 3\times10^{23} R_{14}^{-2} {\rm cm^{-2}}.
\end{equation}
Even with large uncertainties, this is still a substantial column
depth with the promise of severely obscuring any soft X-ray emission.

\section{Emergent Flux}

If the soft X-ray flux generated by SN~Ia progenitors is absorbed, it
will appear at other wavelengths. One estimate of this process is
to assume that the X-ray flux is thermalized and radiated as a black
body at another appropriate temperature. The SSS have characteristic
luminosities around the Eddington limit for a solar mass, $\sim 10^{38}$
erg s$^{-1}$. Let us assume for the sake of argument, that the soft
X-rays are absorbed and re-emitted at a characteristic radius of 
$10^{12} R_{12}$ cm. The characteristic temperature of the re-emitted
radiation is then about 
\begin{equation}
\label{Teff}
T_{\rm eff} \sim 1.9\times10^4 L_{38}^{1/4} R_{12}^{-1/2}~ {\rm K}.
\end{equation}
For this choice of parameters, the frequency at the peak of the
black body distribution would be, by Wien's law, about 
\begin{equation}
\label{lambda}
\lambda \sim 250 L_{38}^{-1/4} R_{12}^{1/2} ~{\rm nm}.
\end{equation}

With the approximation that $\lambda L_{\lambda} \sim L$ and
using Eq. \ref{lambda} for the characteristic wavelength, 
we can write the unextincted flux density as 
\begin{equation}
\label{flux0}
f_{\lambda} \sim \frac{L}{4\pi \lambda D^2}. 
\end{equation}
With $D_{\rm pc}$ the distance to the 
source in parsecs, the flux density can be written as,
\begin{equation}
\label{flux}
f_{\lambda} \sim 3\times10^{-4} L_{38}^{5/4} R_{12}^{-1/2} D_{\rm pc}^{-2}
~{\rm erg~s}^{-1}~{\rm \AA}^{-1}~{\rm cm^{-2}}.
\end{equation} 

\begin{figure}[htp]
\centering
\includegraphics[width=\textwidth]{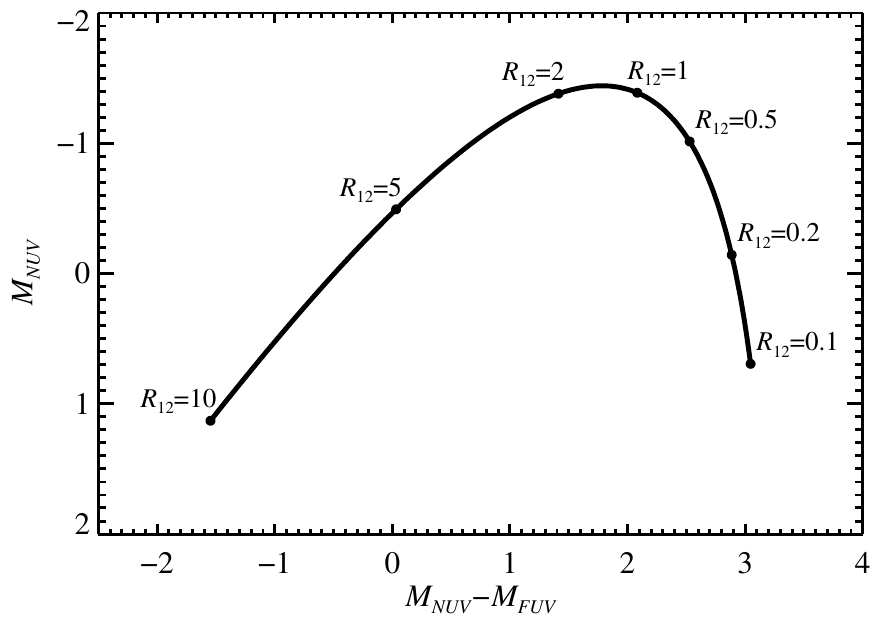}
\figcaption[uvcmd.ps]
{The color magnitude diagram for FUV and NUV flux is given
for black bodies of specified radius and luminosity. 
\label{cmd}}
\end{figure}

For typical radii $\sim10^{12}$ cm, the re-radiated flux would be emitted in 
the UV and would itself be subject to extinction in the in the circumbinary 
material and the ISM. As a simple illustration of one possibility, we assumed 
a fiducial luminosity of $10^{38}$~\ergsec\ and that this luminosity was 
re-radiated from a perfect blackbody of various radii. These blackbody spectra 
were then passed through the \galex\ NUV and FUV responses. The emergent fluxes 
were converted to magnitudes according the \galex\ prescriptions\footnote{http://galexgi.gsfc.nasa.gov/docs/galex/FAQ/counts\_background.html}.  
A distance of 10 pc was assumed in order to compute absolute magnitudes.  
The results are shown in Figure~\ref{cmd}.  As one moves to large radii, the 
blackbody temperature drops, and most of the flux is well outside either \galex\ 
bandpass where analogous limits to those we derive here would apply.

\begin{figure}[htp]
\centering
\includegraphics[width=\textwidth]{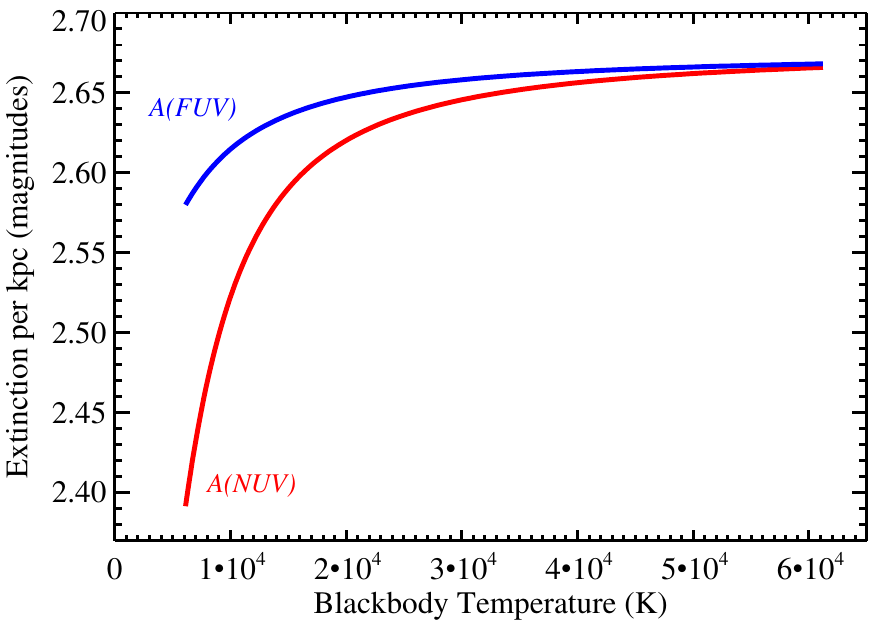}
\figcaption[extinction.pdf]
{The NUV and FUV extinction per kpc based on a blackbody spectral model as a function of the temperature of the blackbody.. 
\label{extinct}}
\end{figure}

The re-radiated flux will be subject to circumbinary and interstellar absorption.  
Here we ignore the former in order to estimate a lower limit to the UV extinction and
hence an upper limit to the number of sources that might be detected in the UV.
To compute these effects, the FUV and NUV magnitudes were calculated without 
any extinction and with the extinction law given by Cardelli et al.\ (1989).  The 
absorption effects depend both on the shape of the spectrum in each band (i.e., 
the temperature of the re-radiated blackbody) and the amount of interstellar 
material the light passes through (i.e., the distance to the source).  We took the 
extinction in the optical to be 1 magnitude per kpc, and we calculated the absorption 
of all the models. The extinction does depend on temperature, as shown in 
Figure~\ref{extinct}.  We take as representative values $A(NUV) = A(FUV) = 
2.65~D_{\rm kpc}$, where $D_{\rm kpc}$ is the distance in kpc.  We note that
Rey et al. (2007) addressed the extinction at a single wavelength corresponding 
to the characteristic wavelength of the NUV pass band. Our procedure of
integrating over the pass band is somewhat more accurate and somewhat
more conservative, yielding a slightly smaller extinction. The absolute magnitudes 
shown in Figure~\ref{cmd} can be readily translated to a particular distance using 
these $A(NUV)$ and $A(FUV)$ relations and a distance modulus.

\begin{figure}[htp]
\centering
\includegraphics[width=\textwidth]{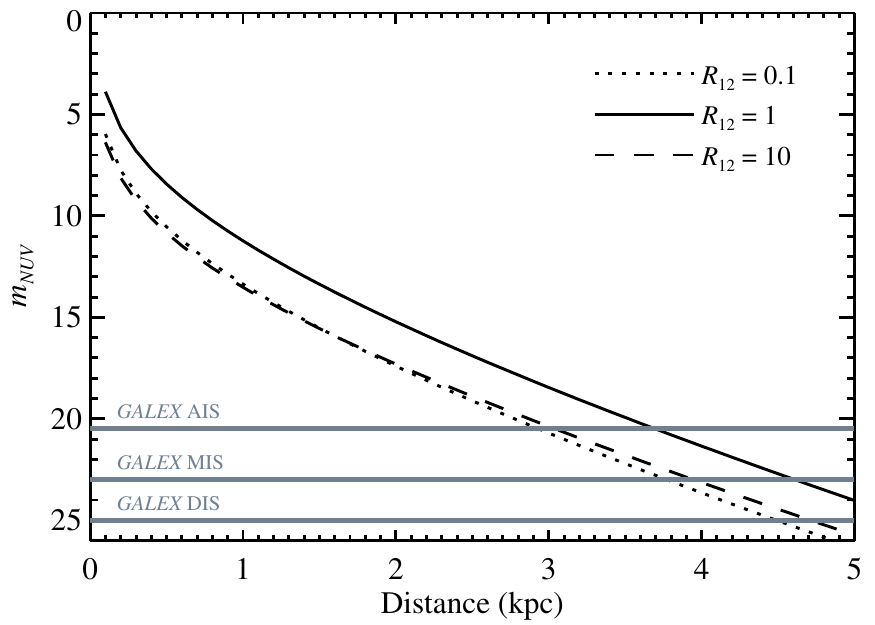}
\figcaption[galex_limits.pdf]
{The apparent NUV magnitude ($m_{\it NUV}$) is shown as a function of distance, taking extinction into account, for each fiducial luminosity and three values of $R_{12}$.  Also shown are the limiting magnitudes for three large-area \galex\ surveys: the All-sky Survey (AIS) covered an area of 40,000 deg$^2$; the Medium Imaging Survey (MIS) covered 1,000 deg$^2$; and the Deep Imaging Survey (DIS) covered only 80 deg$^2$. 
\label{limits}}
\end{figure}

We have taken both distance and $A(NUV)$ into account in Figure~\ref{limits}, 
which shows the apparent NUV magnitude ($m_{\it NUV}$) as a function of distance 
for the fiducial luminosity and three values of $R_{12}$.  Also shown are the limiting 
magnitudes for three large-area \galex\ surveys.  The All-sky Survey (AIS) covered an 
area of 40,000 deg$^2$ and reached a limiting magnitude of 20.5.  The Medium 
Imaging Survey (MIS) covered 1,000 deg$^2$ and reached a limiting magnitude of 23.  
The Deep Imaging Survey (DIS) covered only 80 deg$^2$ and reached a limit of 25 magnitudes.

For all but the lowest temperature blackbodies (those re-radiated at large radii), the
re-radiated emission is brighter than the detection limits of the \galex\ All-sky Survey.
Such black--body re-radiating sources and should be detectable out to $\sim$3.7~kpc
for $R_{12} = 1$ and to $\sim$3.0~kpc for $R_{12}$ = 0.1 or 10. An area around the 
Sun of this radius represents about 10\% of the area of the optical disk of the
Galaxy ($\sim 10^9$ pc$^2$). Di Stefano (2010a) estimates that if the 
SD scenario provides the bulk of SN~Ia, the Galaxy should contain $\sim$ 
1000 nuclear shell-burning white dwarfs that will explode within $10^5$ 
years. If the soft X-ray flux from those sources is absorbed and 
re-emitted in the UV as we have assumed here, then there should be $\sim 100$ bright 
UV sources within the limits of the \galex\ All-sky Survey. The DIS goes much
fainter and could, in principle detect a source to 5 kpc, but examined a much 
smaller portion of the sky, so one would expect at most a single source to
be detected. Searches for such bright UV sources would be of interest. If there remains 
circumbinary matter that can absorb UV, then the absorbed X-ray luminosity would 
come out in yet another band pass. We leave to others to do a proper estimation of 
the re-radiated flux in more general and realistic situations.

\section{Conclusions}   

Small amounts of CSM matter local to the progenitor binary
system of SN~Ia could easily suppress X-ray emission from 
nuclear burning on the surface of an accreting white dwarf.
This suggests that the paucity of SSS to account for
the required rate of explosion of SN~Ia in either the SD
or DD models is not beyond understanding in terms of local
extinction.

Gaining some perspective on the role of SSS in the production
of SN~Ia does not address all the issues associated with
understanding the progenitors of SN Ia. In the SD model, 
the mass transfer rate is required to be sufficiently high
to avoid degenerate, unstable shell ignition and explosion
on the surface of the white dwarf. Published models that
satisfy this constraint and also provide a reasonable number
of progenitor systems, locally extincted or not, require
the mass-transferring secondary star to be a moderately massive
main sequence star ($> 1.16$ \msun; Schaefer \& Pagnotta 2012)
a sub-giant or giant star. The recent advent of SN 2011fe,
an apparently normal ``plain vanilla" SN~Ia, has provided new 
constraints on the progenitor systems. Nugent et al. (2012)
argue that lack of light-curve contamination implies that
the secondary star was not a red giant, and more likely to
be a main sequence star. Li et al. (2012) use archival images
to put limits on the companion and rule out luminous red giants
and almost all helium star models. Bloom et al. (2012)
show that the exploding star was a white dwarf, as expected, 
and that the secondary star was likely to have had a 
radius less than 0.1 that of the Sun, excluding companion
red-giant and main-sequence stars that fill their Roche lobes. 

SNR 0509-67.5 in the LMC was established by scattered, 
time-delayed spectra to be a SN~Ia of the SN~1991T spectral
subclass that exploded about 400 years ago (Rest et al. 2008). 
Schaefer \& Pagnota examined deep HST images of this remnant
to put even tighter limits on the progenitor of this explosion. 
They found that any secondary star must be dimmer than $M_V \sim
8.4$ mag, ruling out basically all published SD models, including 
those with companion main sequence stars of greater than about 
1 \msun, sub-giants, giants, and those involving the stripped 
cores of evolved stars. While one might adopt the dodge that
this was a single event responsible for a somewhat peculiar
and ill-understood sub-class of SN~Ia, and hence not typical
of ``plain vanilla" SN~Ia, these limits remain
a very tight constraint on SD models. Either SD models must
be rejected for this system, or some means must be found to 
impeach the current set of SD models, virtually all of which 
are based on one-dimensional, spherically-symmetric, non-rotating, 
non-magnetic accretion that is undoubtedly incorrect, at least
in detail.

The bulk of this note was written in January, 2012, independent of
the recent posting by Nielsen et al. (2012), who make some of
the same points from a different perspective.

\acknowledgments

We thank Brad Schaefer, Ashley Pagnotta, Rosanne DiStefano, 
and Rob Robinson for constructive discussions.
This research is supported in part by NSF AST-1109801.

\newpage

\newpage

\newpage

\newpage

\newpage

\begin{thebibliography}{}

\bibitem[Badenes et al.(2007)]{2007ApJ...662..472B} 
Badenes, C., Hughes, J.~P., Bravo, E., \& Langer, N.\ 2007, \apj, 662, 472

\bibitem[Blondin et al.(2009)]{2009ApJ...693..207B} 
Blondin, S., Prieto, J.~L., Patat, F., et al.\ 2009, \apj, 693, 207 

\bibitem[Bloom et al.(2012)]{2012ApJ...744L..17B} 
Bloom, J.~S., Kasen, D., Shen, K.~J., et al.\ 2012, \apjl, 744, L17

\bibitem[Cardelli et al.(1989)]{1989ApJ...345..245C} Cardelli, J.~A., 
Clayton, G.~C., \& Mathis, J.~S.\ 1989, \apj, 345, 245 

\bibitem[Di Stefano(2010)]{2010ApJ...712..728D} 
Di Stefano, R.\ 2010a, \apj, 712, 728 

\bibitem[Di Stefano(2010)]{2010ApJ...719..474D} 
Di Stefano, R.\ 2010b, \apj, 719, 474 

\bibitem[Fujimoto(1982)]{1982ApJ...257..767F} 
Fujimoto, M.~Y.\ 1982, \apj, 257, 767 

\bibitem[Gerardy et al.(2004)]{2004ApJ...607..391G} 
Gerardy, C.~L., H{\"o}flich, P., Fesen, R.~A., et al.\ 2004, \apj, 607, 391 

\bibitem[Hachisu et al.(1996)]{1996ApJ...470L..97H} 
Hachisu, I., Kato, M., \& Nomoto, K.\ 1996, \apjl, 470, L97

\bibitem[Hachisu et al.(2010)]{2010ApJ...724L.212H} 
Hachisu, I., Kato, M., \& Nomoto, K.\ 2010, \apjl, 724, L212 

\bibitem[Hamuy et al.(2003)]{2003Natur.424..651H} 
Hamuy, M., Phillips, M.~M., Suntzeff, N.~B., et al.\ 2003, \nat, 424, 651 

\bibitem[Hayden et al.(2010)]{2010ApJ...722.1691H} Hayden, B.~T., 
Garnavich, P.~M., Kasen, D., et al.\ 2010, \apj, 722, 1691

\bibitem[Iben(1982)]{1982ApJ...259..244I} 
Iben, I., Jr.\ 1982, \apj, 259, 244

\bibitem[Iben \& Tutukov(1984)]{1984ApJS...54..335I} 
Iben, I., Jr., \& Tutukov, A.~V.\ 1984, \apjs, 54, 335 

\bibitem[Kahabka \& van den Heuvel(1997)]{1997ARA&A..35...69K} 
Kahabka, P., \& van den Heuvel, E.~P.~J.\ 1997, \araa, 35, 69

\bibitem[Kasen(2010)]{2010ApJ...708.1025K} Kasen, D.\ 2010, \apj, 708, 1025 

\bibitem[Kato \& Hachisu(1999)]{1999ApJ...513L..41K} 
Kato, M., \& Hachisu, I.\ 1999, \apjl, 513, L41 

\bibitem[Mazzali et al.(2005)]{2005ApJ...623L..37M} 
Mazzali, P.~A., Benetti, S., Altavilla, G., et al.\ 2005, \apjl, 623, L37

\bibitem[Neilsen et al.(2012)]{2012Nielsen}Nielsen, M. T. B., Dominik, C., Nelemans, G. \& Voss, R. 2012,
\aap, in press (arXiv:1207.6310)

\bibitem[Nomoto(1982)]{1982ApJ...253..798N} 
Nomoto, K.\ 1982, \apj, 253, 798 

\bibitem[Nugent et al.(2011)]{2011Natur.480..344N} 
Nugent, P.~E., Sullivan, M., Cenko, S.~B., et al.\ 2011, \nat, 480, 344 

\bibitem[Patat et al.(2009)]{2009A&A...508..229P} 
Patat, F., Baade, D., H{\"o}flich, P., et al.\ 2009, \aap, 508, 229 

\bibitem[Patat(2011)]{2011arXiv1109.5799P} Patat, F.\ 2011, arXiv:1109.5799 

\bibitem[Patat et al.(2011)]{2011A&A...530A..63P} 
Patat, F., Chugai, N.~N., Podsiadlowski, P., et al.\ 2011, \aap, 530, A63 

\bibitem[Rest et al.(2008)]{2008ApJ...680.1137R} 
Rest, A., Matheson, T., Blondin, S., et al.\ 2008, \apj, 680, 1137

\bibitem[Rey et al.(2007)]{2007ApJS..173..643R} 
Rey, S.-C., Rich, R.~M., Sohn, S.~T., et al.\ 2007, \apjs, 173, 643

\bibitem[Schaefer(2010)]{2010ApJS..187..275S} 
Schaefer, B.~E.\ 2010, \apjs, 187, 275 

\bibitem[Schaefer \& Pagnotta(2012)]{2012Natur.481..164S} 
Schaefer, B.~E., \& Pagnotta, A.\ 2012, \nat, 481, 164 

\bibitem[Shen \& Bildsten(2008)]{2008ApJ...678.1530S} 
Shen, K.~J., \& Bildsten, L.\ 2008, \apj, 678, 1530 

\bibitem[Simon et al.(2009)]{2009ApJ...702.1157S} 
Simon, J.~D., Gal-Yam, A., Gnat, O., et al.\ 2009, \apj, 702, 1157

\bibitem[van den Heuvel et al.(1992)]{1992A&A...262...97V} 
van den Heuvel, E.~P.~J., Bhattacharya, D., Nomoto, K., \& 
Rappaport, S.~A.\ 1992, \aap, 262, 97 

\bibitem[Wang et al.(2003)]{2003ApJ...591.1110W} 
Wang, L., Baade, D., H{\"o}flich, P., et al.\ 2003, \apj, 591, 1110 

\bibitem[Wang \& Wheeler(2008)]{2008ARA&A..46..433W} 
Wang, L., \& Wheeler, J.~C.\ 2008, \araa, 46, 433 

\bibitem[Webbink(1984)]{1984ApJ...277..355W} 
Webbink, R.~F.\ 1984, \apj, 277, 355 

\bibitem[Wheeler \& Hansen(1971)]{1971Ap&SS..11..373W} 
Wheeler, J.~C., \& Hansen, C.~J.\ 1971, \apss, 11, 373 

\bibitem[Whelan \& Iben(1973)]{1973ApJ...186.1007W} 
Whelan, J., \& Iben, I., Jr.\ 1973, \apj, 186, 1007 

\end{thebibliography}
\end{document}